\documentclass[letterpaper,preprintnumbers,twocolumn,superscriptaddress,aps,nofootinbib,perprint,prl]{revtex4}
\usepackage{amssymb}
\usepackage[centertags]{amsmath}
\usepackage{txfonts}
\usepackage{epsfig}
\usepackage{bm}
\usepackage{color}
\usepackage{graphicx,graphics}
\usepackage{multirow}
\usepackage{float}
\usepackage[pdfstartview=FitH]{hyperref}
\hypersetup{colorlinks=true, citecolor=blue, linkcolor=blue,filecolor=black,urlcolor=blue}
\allowdisplaybreaks[2]

\usepackage{slashed}
\usepackage{ulem}

\usepackage{booktabs}
\usepackage{multirow}

\begin{document}

\title{Twist-4 contributions to semi-inclusive deeply inelastic scatterings with polarized beam and target}

\author{Shu-yi Wei}
\affiliation{Key Laboratory of Quark and Lepton Physics (MOE) 
  and Institute of Particle Physics,  Central China Normal University,  Wuhan 430079, China}

\author{Yu-kun Song}
\affiliation{School of Physics and Technology,  University of Jinan,   Jinan, Shandong 250022, China}

\author{Kai-bao Chen}
\affiliation{School of Physics \&  Key Laboratory of Particle Physics and Particle Irradiation (MOE), 
  Shandong University,  Jinan, Shandong 250100, China }

\author{Zuo-tang Liang}
\affiliation{School of Physics \&  Key Laboratory of Particle Physics and Particle Irradiation (MOE), 
  Shandong University,   Jinan, Shandong 250100, China }

\begin{abstract}
We present for the first time the complete twist-4 result for the semi-inclusive deeply inelastic scattering 
$e^- N \to e^-qX$ with polarized electron and proton beams at the tree level of pQCD.  
The calculations have been carried out using the formulism obtained after collinear expansion 
where the multiple gluon scattering are taken into account and gauge links are obtained automatically in a systematical way. 
The results show in particular that there are twist-4 contributions to all the eight twist-2 structure functions 
for $e^- N \to e^-hX$ that correspond to the eight twist-2 transverse momentum dependent parton distribution functions. 
Such higher twist effects could be very significant and 
therefore have important impacts on extracting these three-dimensional parton distribution functions 
from the asymmetry data on $e^- N \to e^-hX$.  
We suggest also an approximate way for a rough estimation of such higher twist contributions. 
\end{abstract}


\maketitle

{\it{Introduction}} ---
Three-dimensional or transverse momentum dependent (TMD) parton distribution functions (PDFs) 
are one of the frontiers in hadron physics in particular in the study of hadron structure 
and properties of quantum chromodynamics (QCD)~\cite{Anselmino2016,Liang:2015nia}. 
When the transverse momentum of the parton is concerned, the sensitive measurable quantities 
in high energy reactions are often different azimuthal angle asymmetries. 
Higher twist contributions could be very significant thus play an important role 
when going from the one-dimensional to the three-dimensional case. 

In contrast to twist-3 contributions that often lead to azimuthal asymmetries that are missing 
when only twist-2 contributions are considered (see e.g. \cite{Mulders:1995dh,Bacchetta:2006tn,Boer:1997mf}), 
in many cases, twist-4 contributions are just addenda to twist-2 asymmetries. 
Since the asymmetries themselves are usually not very large, twist-4 contributions can be relatively very significant 
and have large influences on determining the twist-2 PDFs from experimental data. 
This is particularly the case in view that most of the data currently available are from experiments 
at not very high energies (see e.g. \cite{Airapetian:1999tv,Airapetian:2009ae,Ageev:2006da,Avakian:2010ae,Qian:2011py,Huang:2011bc}, or \cite{Barone:2010zz} for a recent review).
It is therefore necessary and important to make systematic studies including higher twist contributions. 
However, such a systematic study up to twist-4 is very complicated and might be even impossible  
in particular because we need to deal with quark-2-gluon-quark correlators with three independent parton momenta and 
other related complicated problems.   

Higher twist effects in inclusive deeply inelastic lepton-nucleon scattering (DIS) and Drell-Yan processes
have been studied already in 1980s to 90s~\cite{Ellis:1982wd, Qiu:1990xy}.
It has been shown that the collinear expansion is a necessary procedure for obtaining the hadronic tensor or 
the cross section in terms of gauge invariant one-dimensional PDFs. 
More recently, it has been shown that~\cite{Liang:2006wp} collinear expansion 
can be extended to the semi-inclusive DIS process $e^- N \to e^- qX$, 
where $q$ denotes a quark that corresponds to a jet of hadrons in experiments.
In the formulism obtained, the multiple gluon scattering are taken into account and gauge links are obtained automatically and systematically. 
Moreover, the expressions for the hadronic tensor obtained after the collinear expansion are simple and elegant 
in the sense that they are given in terms of PDFs and hard parts. 
The hard parts are not only calculable but also simplified to a form independent of the parton momentum,  
and correspondingly the involved PDFs are not only gauge invariant but also all defined via quark-quark or quark-$j$-gluon-quark correlators 
with only one independent parton momentum left.  
This makes the expressions much simpler and higher twist calculations are more feasible. 
Based on this formulism,  the complete twist-3 result for $e^- N \to e^- qX$ and $e^+ e^- \to hqX$  
have been obtained and are presented in \cite{Liang:2006wp,Song:2013sja,Wei:2013csa}.   
Although there are still large differences between $e^- N \to e^- qX$ and $e^- N \to e^- hX$, 
the study of the former can provide useful references at least qualitatively for the latter. 

In this paper, we present for the first time the complete twist-4 result for $e^- N \to e^- qX$ with polarized electron and nucleon beams. 
After a brief summary of the general formulism, we calculate the structure functions at the tree level of pQCD up to twist-4. 
We present also the results for azimuthal asymmetries and suggest an approximate method for a rough estimation of twist-4 contributions. 

{\it{The formulism}} --- 
To be explicit, we consider the semi-inclusive DIS (SIDIS) $e^-N\to e^-qX$.  
The cross section is given in terms of the well-known leptonic tensor  $L^{\mu\nu}$ and the hadronic tensor $W_{\mu\nu}$ as,
\begin{align}
d\sigma = \frac{\alpha^2_{\rm em}e_q^2}{sQ^4} 
L^{\mu\nu}(l,\lambda_l,l')W_{\mu\nu}(q,p,S,k) \frac{d^3l'd^3k}{(2\pi)^32E_{l'}2E_{k}},
\end{align}
where $l$, $l'$, $p$ and $k$ are the 4-momenta of the incident, the outgoing electron, the incident nucleon and the outgoing quark $q$ respectively;  
$q=l-l'$ and $Q^2=-q^2$; $\lambda_l$ and $S$ are the helicity of the electron and the spin of the nucleon. 
%

A general kinematic analysis shows that the cross section is given in terms of 18 structure functions describing different polarization cases and/or 
azimuthal asymmetries (see e.g. \cite{Bacchetta:2006tn}).  
In the QCD parton model, after collinear expansion, the hadronic tensor is expressed in terms of 
the gauge invariant quark-quark and quark-$j$-gluon-quark correlators and calculable hard parts~\cite{Liang:2006wp,Song:2013sja},
$W_{\mu\nu}=\sum_{j,c}\tilde W_{\mu\nu}^{(j,c)}$,
where $j$ denotes the number of gluons exchanged and $c$ denotes eventually different cuts. 
After integration over $k^-$, these $\tilde W_{\mu\nu}$'s are simplified to~\cite{Song:2013sja}, 
\begin{align}
  & \tilde W_{\mu\nu}^{(0)} =  {\rm Tr}[\hat h_{\mu\nu}^{(0)} \hat \Phi^{(0)}]/2, \label{eq:W0}   \\
  & \tilde W_{\mu\nu}^{(1,L)} ={\rm Tr} [\hat h_{\mu\nu}^{(1)\rho} \hat \varphi^{(1)}_{\rho}] /{2p\cdot q} ,   \label{eq:W1L}  \\
  & \tilde W_{\mu\nu}^{(2,L)} = {\rm Tr}[\hat N_{\mu\nu}^{(2)\rho\sigma} \hat \varphi_{\rho\sigma}^{(2)}]/{4(p\cdot q)^2},   \label{eq:W2L}  \\
  & \tilde W_{\mu\nu}^{(2,M)} = {\rm Tr}[ \hat h_{\mu\nu}^{(2)\rho\sigma} \hat \varphi_{\rho\sigma}^{(2,M)}]/{4(p\cdot q)^2},   \label{eq:W2M}
\end{align}
where the hard parts are reduced to the simple forms independent of the parton momentum,  i.e.,  
$\hat h_{\mu\nu}^{(0)} =  \gamma_\mu \slashed n \gamma_\nu/{p^+},$
$\hat h_{\mu\nu}^{(1)\rho} = \gamma_\mu \slashed {\bar n} \gamma^\rho_\perp \slashed n \gamma_\nu, $
$\hat N_{\mu\nu}^{(2)\rho\sigma} = q^- \gamma_\mu \gamma_\perp^\rho \slashed n \gamma_\perp^\sigma \gamma_\nu,$
$ \hat h_{\mu\nu}^{(2)\rho\sigma} =    \gamma_\mu \gamma_\perp^\rho \slashed p \gamma_\perp^\sigma  \gamma_\nu,$
$\gamma_\perp^\rho\equiv g_\perp^{\rho\sigma}\gamma_\sigma$ 
and $g_\perp^{\rho\sigma}\equiv g^{\rho\sigma}-\bar n^\rho n^\sigma-\bar n^\sigma n^\rho$.
The involved quark-quark and quark-$j$-gluon-quark correlators are given by, 
\begin{align}
&\hat\Phi^{(0)} (x, k_\perp) =   \int \frac{p^+ dy^- d^2 y_\perp}{(2\pi)^3} e^{ixp^+y^- - i\vec{k}_\perp\cdot \vec{y}_\perp} \nonumber\\
&\phantom{XXXXXXXX} \times  \langle N | \bar \psi (0) \mathcal{L} (0,y) \psi (y) | N \rangle, \label{eq:hPhi0}\\
&\hat\varphi^{(1)}_\rho  (x,  k_\perp) =  \int \frac{p^+ dy^- d^2 y_\perp}{(2\pi)^3}   e^{ixp^+y^- -i\vec{k}_\perp \cdot \vec{y}_\perp} \nonumber\\
&\phantom{XXXXXXXX} \times  \langle N | \bar \psi (0) D_{\perp\rho} (0) \mathcal{L} (0, y) \psi (y) | N \rangle,\label{eq:hphi1}\\
&\hat \varphi_{\rho\sigma}^{(2)}  (x,  k_\perp)  =  \int \frac{p^+dy^-d^2y_\perp}{(2\pi)^3} \int_0^\infty ip^+dz^-  e^{ixp^+y^--i\vec{k}_\perp\cdot\vec{y}_\perp} \nonumber\\
&\phantom{XXXXX} \times \langle N | \bar \psi(0) \mathcal{L}(0,z) D_{\perp\rho}(z) D_{\perp\sigma} (z) \mathcal{L}(z,y) \psi(y) |N\rangle,\label{eq:hphi2}\\
& \hat \varphi_{\rho\sigma}^{(2,M)} (x,  k_\perp) =  \int \frac{p^+dy^- d^2y_\perp}{(2\pi)^3} e^{ixp^+y^--i\vec{k}_\perp\cdot\vec{y}_\perp}\nonumber\\
&\phantom{XXXXXXXX} \times \langle N | \bar\psi(0) D_{\perp\rho} (0) \mathcal{L} (0,y) D_{\perp\sigma} (y) \psi (y) | N \rangle,\label{eq:hphi2M}
\end{align}
where $D_\rho=-i\partial_\rho +g A_\rho$, and ${\cal L}(0,y)$ is the well-known gauge link. 
We see that the involved $\hat\varphi^{(j)}$ are all $D$-type and are simplified to depend only on one independent parton momentum. 

In $e^-N\to e^-qX$, where the fragmentation is not considered, only chiral even PDFs  are involved. 
We need only the $\gamma^\alpha$- and $\gamma_5\gamma^\alpha$-terms in the expansion
of the correlator in terms of the $\Gamma$-matrices, e.g.,  
$\hat\Phi^{(0)}   = ( \Phi^{(0)}_\alpha \gamma^\alpha - \tilde \Phi^{(0)}_\alpha \gamma_5\gamma^\alpha+ \cdots)/2$. 

Up to twist-4, we need the complete Lorentz decompositions of $\Phi^{(0)}_\alpha$ and $\tilde\Phi^{(0)}_\alpha$.
The twist-4 parts are given by~\cite{Goeke:2005hb},
\begin{align}
  \Phi^{(0)}_{3\alpha}=
  &\frac{M^2}{p^+}n_\alpha\Bigl(f_3 - \frac{\varepsilon_\perp^{kS}}{M}f_{3T}^\perp\Bigr), \label{eq:phi0lorentzdecom} \\
 \tilde\Phi^{(0)}_{3\alpha}  = 
 & -\frac{M^2}{p^+} n_\alpha   \Bigl( \lambda_h g_{3L}  - \frac{k_\perp\cdot S_T}{M}  g_{3T}^\perp  \Bigr), \label{eq:tphi0lorentzdecom}
\end{align}
where $\varepsilon_\perp^{kS}=\varepsilon_{\perp\alpha\beta}k_\perp^\alpha S_T^\beta$ and $\varepsilon_{\perp\alpha\beta}=\varepsilon_{\alpha\beta\rho\sigma}\bar n^\rho n^\sigma$. 
Here, we use the naming system as that used in \cite{Bacchetta:2006tn,Song:2013sja,Wei:2013csa,Goeke:2005hb,Chen:2016moq}, 
where $f$'s and $g$'s are defined from the $\gamma_\alpha$- and $\gamma_5 \gamma_\alpha$-term respectively;  
a digit $j$ in the subscript stands for twist-$(j+1)$, those without $j$ are for twist-3; 
the subscript $T$ or $L$ denotes hadron polarization, those without $T$ or $L$ denote unpolarized. 
A subscript $3$ is added to $\Phi$ and $\tilde\Phi$ to denote the twist-4 parts only.
We see that there are 4 chiral even twist-4 TMD PDFs defined via $\hat\Phi^{(0)}$.  
For $\hat\varphi^{(1)}$ and $\hat\varphi^{(2)}$, the chiral even twist-4 parts are, 
\begin{align}
  \varphi^{(1)}_{3\rho\alpha} =
  & M^2 g_{\perp\rho\alpha} \Bigl(f_{3d}- \frac{\varepsilon_\perp^{kS}}{M} f_{3dT}^\perp\Bigr) \nonumber\\    
  & + \Bigl(k_{\perp\rho}k_{\perp\sigma}-\frac{k_\perp^2}{2} g_{\perp\rho\sigma}\Bigr) \Bigl(f_{3d}^\perp+\frac{\varepsilon_\perp^{kS}}{M}f_{3dT}^{\perp 2}\Bigr) \nonumber \\ 
  & + i M^2\varepsilon_{\perp\rho\alpha} \Bigl(\lambda_h f_{3dL}-\frac{k_\perp\cdot S_T}{M} f_{3dT}^{\perp 3}\Bigr) \nonumber\\   
  & + (k_{\perp\{\rho} \tilde k_{\perp \alpha \}}/2) \Bigl(\lambda_h f_{3dL}^\perp+\frac{k_\perp\cdot S_T}{M}f_{3dT}^{\perp 4}\Bigr), \label{eq:phi1lorentzdecom} \\
%
  \tilde\varphi^{(1)}_{3\rho\alpha} = 
  & iM^2\varepsilon_{\perp\rho\alpha} \Bigl(g_{3d} - \frac{\varepsilon_\perp^{kS}}{M} g_{3dT}^\perp\Bigr) \nonumber\\   
  &  + i (k_{\perp\{\rho} \tilde k_{\perp \alpha \}}/2) \Bigl(g_{3d}^\perp + \frac{\varepsilon_\perp^{kS}}{M}  g_{3dT}^{\perp 2}\Bigr) \nonumber\\ 
  & +M^2 g_{\perp\rho\alpha}\Bigl(\lambda_h g_{3dL} - \frac{k_\perp\cdot S_T}{M} g_{3dT}^{\perp 3}\Bigr) \nonumber\\ 
  & + i  \Bigl(k_{\perp\rho}k_{\perp\alpha}-\frac{k_\perp^2}{2} g_{\perp\rho\alpha}\Bigr) \Bigl(\lambda_h g_{3dL}^\perp+\frac{k_\perp\cdot S_T}{M} g_{3dT}^{\perp 4}\Bigr), \\
%
 \varphi_{3\rho\sigma\alpha}^{(2)}  = & p^+\bar n_\alpha \Bigl[
   M^2 g_{\perp\rho\sigma} \Bigl(f_{3dd} - \frac{\varepsilon_\perp^{kS} }{M} f_{3ddT}^{\perp}\Bigr)     \nonumber\\ 
   & + \Bigl(k_{\perp\rho} k_{\perp\sigma} - \frac{k_\perp^2}{2} g_{\perp\rho\sigma}\Bigr) 
   \Bigl(f_{3dd}^{\perp} + \frac{\varepsilon_\perp^{kS}}{M}  f_{3ddT}^{\perp 2}\Bigr)  \nonumber \\ 
& + iM^2\varepsilon_{\perp\rho\sigma} \Bigl(\lambda_h f_{3ddL} - \frac{k_\perp\cdot S_T}{M} f_{3ddT}^{\perp 3}\Bigr)  \nonumber\\ 
& + (k_{\perp\{\rho} \tilde k_{\perp \sigma \} }/2)  \Bigl(\lambda_h f_{3ddL}^{\perp}+\frac{k_\perp \cdot S_T}{M} f_{3ddT}^{\perp 4}\Bigr)  \Bigr], \label{eq:phi2lorentzdecom} \\
 \tilde\varphi_{3\rho\sigma\alpha}^{(2)}  =& p^+\bar n_\alpha \Bigl[ 
    iM^2 \varepsilon_{\perp\rho\sigma} \Bigl(g_{3dd} - \frac{\varepsilon_\perp^{kS}}{M}  g_{3ddT}^{\perp}\Bigr)  \nonumber\\ 
 & + (k_{\perp\{\rho} \tilde k_{\perp \sigma \} }/2) \Bigl(g_{3dd}^{\perp} + \frac{\varepsilon_\perp^{kS}}{M} g_{3ddT}^{\perp 2}\Bigr) \nonumber\\
 & + M^2 g_{\perp\rho\sigma} \Bigl(\lambda_h g_{3ddL} - \frac{k_\perp\cdot S_T}{M} g_{3ddT}^{\perp 3}\Bigr) \nonumber\\ 
 + & \Bigl(k_{\perp\rho} k_{\perp\sigma} - \frac{k_\perp^2}{2} g_{\perp\rho\sigma}\Bigr) \Bigl(\lambda_h g_{3ddL}^{\perp} + \frac{k_\perp\cdot S_T}{M} g_{3ddT}^{\perp 4}\Bigr)  \Bigr], \label{eq:tphi2lorentzdecom} 
\end{align}
where we, as in \cite{Chen:2016moq}, add in the subscript a lower case letter $d$ or a $dd$ to denote 
TMDs defined via the $D$-type quark-gluon-quark or quark-2-gluon-quark correlator; 
$\tilde k_{\perp\alpha}\equiv\varepsilon_{\perp\alpha\beta}k_\perp^\beta$, 
and $k_{\perp\{\rho} \tilde k_{\perp \alpha \}}\equiv k_{\perp\rho} \tilde k_{\perp \alpha} + k_{\perp\alpha} \tilde k_{\perp \rho}$.
Besides the superscript $M$, those for $\hat\varphi^{(2,M)}$ are exactly the same as $\hat\varphi^{(2)}$.  
Since there are more than one $f_{3dT}^\perp$'s and $g_{3dT}^\perp$'s according to the naming rules, 
we introduce an additional digit in the superscript to distinguish them from each other. 
They are similar in decompositions of $\hat\varphi^{(1)}$ and $\hat\varphi^{(2)}$.  
Totally we have 4 $f_{3dT}^\perp$'s associated with the four independent Lorentz tensors $g_{\perp\rho\alpha}$, 
$(k_{\perp\rho}k_{\perp\sigma}-{k_\perp^2}g_{\perp\rho\sigma}/2)$, $\varepsilon_{\perp\rho\alpha}$ and $k_{\perp\{\rho} \tilde k_{\perp \alpha \}}$ respectively; 
while $g_{3dT}^\perp$ to $g_{3dT}^{\perp4}$ are associated with 
$\varepsilon_{\perp\rho\alpha}$, $k_{\perp\{\rho} \tilde k_{\perp \alpha \}}$,
$g_{\perp\rho\alpha}$, $(k_{\perp\rho}k_{\perp\sigma}-{k_\perp^2} g_{\perp\rho\sigma}/2)$ respectively. 
The four Lorentz tensors are orthogonal to each other.

We see that for the twist-4 parts, the decompositions of $\varphi$ and $\tilde\varphi$ have exact one to one correspondence. 
For each $f_{3}$, there is correspondingly a $g_{3}$. They always appear in pair. 
We have totally 8 such pairs from $\hat\varphi^{(1)}$, $\hat\varphi^{(2)}$ and $\hat\varphi^{(2,M)}$ respectively.   
Due to the Hermiticity of $\hat\Phi^{(0)}$ and $\hat\varphi^{(2,M)}$,  PDFs defined via these two correlators are real. 
However, those defined via  $\hat\varphi^{(1)}$ and $\hat\varphi^{(2)}$ are complex containing both real and imaginary parts.

From the QCD equation of motion, $\gamma\cdot D\psi=0$, 
we obtain a series of relationships between the TMD PDFs defined via $\hat\varphi^{(j)}$ and those defined via $\hat\Phi^{(0)}$. 
For the chiral even twist-3 part, they can be given in the unified form~\cite{Chen:2016moq},  
\begin{align}
f_{dS}^K+g_{dS}^K=x(f_{S}^K+ig_{S}^K), \label{eq:t3eom}
\end{align} 
where $S=$null, $L$ or $T$ and $K=$null or $\perp$ whenever applicable. 
Using the relationships given by Eq.~(\ref{eq:t3eom}), 
we can replace all the TMD PDFs defined via $\hat\varphi^{(1)}$ by those defined via $\hat\Phi^{(0)}$ in the 
final twist-3 results for the hadronic tensor in SIDIS~\cite{Bacchetta:2006tn,Song:2013sja}, 
and similar for $e^+e^-$-annihilations~\cite{Wei:2013csa,Chen:2016moq}.    
Similarly, for the chiral even twist-4 part, we obtain, 
\begin{align}
& x^2 f_{3} = x f_{-3d}= - f_{-3dd}^{M}, \ \ \ \ \ \ \ 
x^2 f_{3T}^\perp  = x f_{-3dT}^{\perp}= - f_{-3ddT}^{M\perp}, \label{eq:t4feom}\\
& x^2 g_{3L}  = xf_{-3dL}=-f_{-3ddL}^M, \ \ 
x^2 g_{3T}^\perp = x f_{-3dT}^{\perp 3}= -f_{-3ddT}^{M\perp 3}, \label{eq:t4geom}
\end{align}
where $f_{\pm}\equiv f\pm g$ such as $f_{-3d}\equiv f_{3d}-g_{3d}$ and so on. 
We note that Eqs.~(\ref{eq:t4feom}-\ref{eq:t4geom}) represent 12 real equations and can be used to replace 
those independent twist-4 TMD PDFs in parton model results for cross section.

The operator expressions of these twist-4 TMD PDFs can be obtained by reversing 
the corresponding equations for Lorentz decompositions. 
When the multiple gluon scattering is taken into account, these higher twist TMD PDFs are all new and much involved. 
They reflect not only the parton distributions but also quantum inference effects in the scattering.
There is little data available that gives direct insights on them. 
However, if we neglect the multiple gluon scattering, i.e., put $g=0$,  
we obtain a set of simple equations relating them to the twist-2 counterparts. 
They could be helpful in understanding the significances of these higher twist PDFs in particular at the present stage.

By putting $g=0$ into Eqs.~(\ref{eq:hPhi0}-\ref{eq:hphi2M}), we relate  $\hat\varphi^{(j)}$ to $\hat\Phi^{(0)}$, i.e., 
$\hat\varphi^{(1)}_\rho= - k_{\perp\rho}\hat\Phi^{(0)}$, 
$\hat\varphi^{(2,M)}_{\rho\sigma}=k_{\perp\rho} k_{\perp\sigma}\hat\Phi^{(0)}$, and 
$\hat\varphi^{(2)}_{\rho\sigma}+\hat\varphi^{(2)\dag}_{\sigma\rho}=k_{\perp\rho} k_{\perp\sigma}{\partial}\hat\Phi^{(0)}/{\partial x}$. 
Together with the equation of motion, these relationships relate all higher twist PDFs to leading twist ones. 
For those defined via $\hat\varphi^{(1)}$, we have, 
\begin{align}
& xf_{3d} = \frac{k_\perp^2}{2M^2} xf_{3d}^\perp = x^2f_3 = -\frac{k_\perp^2}{2M^2}f_1,\label{eq:f3dg0} \\
&xg_{3dL} = i\frac{k_\perp^2}{2M^2}xg_{3dL}^\perp = -x^2g_{3L} = \frac{k_\perp^2}{2M^2}g_{1L}, \label{eq:g3dLg0}\\
&xf_{3dT}^{\perp} = -\frac{k_\perp^2}{2M^2} xf_{3dT}^{\perp2} = x^2f_{3T}^\perp = - \frac{k_\perp^2}{2M^2}f_{1T}^\perp,\label{eq:f3dTperpg0} \\
& xg_{3dT}^{\perp3} = - i\frac{k_\perp^2}{2M^2} xg_{3dT}^{\perp4} = -x^2g_{3T}^\perp = \frac{k_\perp^2}{2M^2}g_{1T}^\perp, \label{eq:g3dTperpg0} 
\end{align}
and all the others vanish. For those defined via $\hat\varphi^{(2)}$, we have,  
\begin{align}
&2{\rm Re} f_{3dd} = 2{\rm Re} \frac{k_\perp^2}{2M^2} f_{3dd}^\perp = \frac{k_\perp^2}{2M^2}\frac{\partial}{\partial x} f_1,\label{eq:f3ddg0} \\
&2{\rm Re} g_{3ddL} = 2{\rm Re} \frac{k_\perp^2}{2M^2} g_{3ddL}^\perp = -\frac{k_\perp^2}{2M^2}\frac{\partial}{\partial x} g_{1L},\label{eq:g3ddLg0} \\
&2{\rm Re} f_{3ddT}^{\perp} = -2{\rm Re}\frac{k_\perp^2}{2M^2}f_{3ddT}^{\perp2} = \frac{k_\perp^2}{2M^2}\frac{\partial}{\partial x} f_{1T}^\perp,\label{eq:f3ddTperpg0} \\
&2{\rm Re} g_{3ddT}^{\perp3} = -2{\rm Re}\frac{k_\perp^2}{2M^2}g_{3ddT}^{\perp4} = -\frac{k_\perp^2}{2M^2}\frac{\partial}{\partial x} g_{1T}^\perp,  \label{eq:g3ddTperpg0} 
\end{align}
and all the others vanish. Time reversal invariance demands $f_{1T}^\perp=0$ in this case~\cite{Collins:1992kk}.

{\it{The complete twist-4 result}} --- 
We substitute the Lorentz decompositions of the quark-quark and quark-$j$-gluon-quark correlators 
given by Eqs.~(\ref{eq:phi0lorentzdecom}-\ref{eq:tphi2lorentzdecom}) into Eqs.~(\ref{eq:W0}-\ref{eq:W2M}), 
carry out the calculations, and obtain the hadronic tensor and cross section up to twist-4.  
We compare the results with the general form of the cross section given in e.g. \cite{Bacchetta:2006tn} and obtain the structure functions as, 
\begin{align}
& W_{UU,T} = x  f_1 + {4x^2} \kappa_M f_{+3dd}, \label{eq:wuut} \\
& W_{UU,L} = {8x^3} \kappa_M f_3, \label{eq:wuul} \\
& W_{UU}^{\cos 2\phi} = - 2 x^2\kappa_M\frac{|\vec k_\perp|^2}{M^2} f_{-3d}^{\perp}, \label{eq:wuu3} \\
& W_{UL}^{\sin 2\phi} = 2 x^2\kappa_M \frac{\vec |k_\perp|^2}{M^2} f_{+3dL}^{\perp}, \label{eq:wul} \\
& W_{LL} = x g_{1L} + 4 x^2\kappa_M f_{+3ddL}, \label{eq:wll} \\
& W_{UT,T}^{\sin (\phi-\phi_S)} = \frac{|\vec k_\perp|}{M} 
( x f_{1T}^\perp + 4 x^2\kappa_M f^{\perp}_{+3ddT} ), 
\label{eq:wut2} \\
& W_{UT,L}^{\sin (\phi-\phi_S)} = 8x^3\kappa_M \frac{|\vec k_\perp|}{M}  f_{3T}^\perp, \\
& W_{UT}^{\sin (\phi+\phi_S)} = - x^2 \kappa_M \frac{|\vec k_\perp|^3}{M^3} (f_{+3dT}^{\perp 4} + f_{-3dT}^{\perp 2}) , 
\label{eq:wut3}\\
& W_{UT}^{\sin (3\phi-\phi_S)} = - x^2 \kappa_M \frac{|\vec k_\perp|^3}{M^3} (f_{+3dT}^{\perp 4} - f_{-3dT}^{\perp 2})  
, \label{eq:wut4}\\
& W_{LT}^{\cos (\phi-\phi_S)} = \frac{|\vec k_\perp|}{M} \bigl( xg_{1T}^\perp + {4x^2}\kappa_M f_{+3ddT}^{\perp 3} \bigr), \label{eq:wlt2} 
\end{align}
where $\kappa_M=M^2/Q^2$ symbolizing twist-4 contributions. 
Only the real part of the TMD PDF contributes and this is true for all the twist-4 PDFs involved in Eqs.~(\ref{eq:wuut}-\ref{eq:wlt2})
so we just omit the symbol Re for clarity of the equation.
For the same reason, we omit the overall factor $e^2_q$ and a sum over flavor is also implicit.
To differentiate them from those for $e^-N\to e^-hX$ where fragmentation is also involved, 
we use $W$ with the same sub- and superscripts to represent the counterpart of $F$.  
We present here the 10 structure functions that have twist-4 contributions. 
The other 8 structure functions describe the azimuthal asymmetries given by either the cosine or sine  
of a single $\phi$ or $\phi_S$ or $2\phi-\phi_S$, 
and have only twist-3 contributions up to twist-4. 
These twist-3 results can be found e.g. in \cite{Song:2013sja}.

From the results given by Eqs.~(\ref{eq:wuut})-(\ref{eq:wlt2}), we see clearly the following distinct features:

(1) Up to twist-4, all 18 structure functions are non-zero. 
Besides those 8 that have twist-3 as leading power contributions, all the rest 10 have twist-4 contributions. 
For $e^-N\to e^-qX$ that we consider here,  4 of them have twist-2 and the other 6 have twist-4 as leading power contributions. 
And all the 4 twist-2 structure functions have twist-4 addenda to them. 
It is also very interesting and important to note that the twist-3 part contributes to azimuthal asymmetries 
that are all missing at either twist-2 or twist-4 hence can be studied separately. 
However the twist-4 and twist-2 contributions may mixed up with each other and give rise to the same asymmetry 
hence are difficult to separate them from each other.

(2) We recall that  for $e^-N\to e^-hX$ where fragmentation is considered, 
there are 8 twist-2 structure functions $F$'s that correspond to the 8 twist-2 TMD PDFs. 
We see that all the $W$'s corresponding to them have twist-4 contributions. 
This means that we have to consider twist-4 contributions if we use data on $e^-N\to e^-hX$ to extract the corresponding twist-2 TMD PDFs. 
Since $e^-N\to e^-hX$ is one of the major sources for the data~\cite{Barone:2010zz} 
now available for extracting TMD PDFs, it is thus important to study these twist-4 contributions to get correct and precise knowledge even on twist-2 TMDs. 

(3) If we consider $e^-N\to e^-hX$, besides $W_{UU,L}$ and $W_{UT,L}^{\sin (\phi-\phi_S)}$,  all the twist-4 contributions are addenda to twist-2 structure functions. 
Since $W_{UU,L}$ is added to $W_{UU,T}$ and $W_{UT,L}^{\sin (\phi-\phi_S)}$ to $W_{UT,T}^{\sin (\phi-\phi_S)}$ to give the final observable effects, 
this means that all the twist-4 contributions are addenda to twist-2 contributions in $e^-N\to e^-hX$. 
This makes it very difficult to separate them from each other. 
A clean and perhaps practical way to study twist-4 effects is to study $e^-N\to e^-qX$, i.e. by measuring the jet production. 
In this case we have 6 structure functions that have twist-4 as the leading power contributions and 4 of them correspond to separate azimuthal asymmetries.

Correspondingly, there are two twist-2 azimuthal asymmetries for $e^-N\to e^-qX$ and, up to twist-4, they are given by,  
\begin{align}
&\langle{\sin (\phi-\phi_S)}\rangle_{UT}= \frac{|\vec k_\perp|}{2M}  \frac{f_{1T}^\perp}{f_1}  (1-\alpha_{UT}\kappa_M), \label{eq:Asin1ut} \\
&\langle{\cos (\phi-\phi_S)}\rangle_{LT} =
\frac{ |\vec k_\perp|}{2M} \frac{y(2-y)}{A(y)} \frac{g_{1T}^\perp}{f_1} (1-  \alpha_{LT} \kappa_M), \label{eq:Acoslt2} 
\end{align}
where $A(y)= 1+(1-y)^2$, 
and the twist-4 modification factors are given by, 
\begin{align}
&\alpha_{UT}=\alpha_{UU} - 16x^2 \frac{1-y}{A(y)} \frac{f_{3T}^\perp}{f_{1T}^\perp} - 4x\frac{f_{+3ddT}^{\perp}}{f_{1T}^\perp}, \label{eq:alphaut} \\
%
&\alpha_{LT}=\alpha_{UU} - 4x\frac{f_{+3ddT}^{\perp 3}} {g_{1T}^\perp}, \label{eq:alphalt} 
\end{align}
where $\alpha_{UU}$ is due to twist-4 contributions to $W_{UU}$. 
It is the ratio of the twist-4 to twist-2 contributions in unit of $\kappa_M$, i.e., 
\begin{align}
\alpha_{UU} = 
16x^2\frac{1-y}{A(y)} \frac{f_3}{f_1} + 4x \frac{ f_{+3dd}}{f_1}. \label{eq:alphauu}
\end{align}
There are 4 twist-4 azimuthal asymmetries given by,
\begin{align}
&\langle\cos2\phi\rangle_{UU} = - 2\kappa_M \frac{|\vec k_\perp|^2}{M^2} \frac{1-y} {A(y)}  \frac{x f_{-3d}^{\perp}} {f_1}, \label{eq:Acos2uu} \\
&\langle\sin2\phi\rangle_{UL} = 2\kappa_M \frac{|\vec k_\perp|^2}{M^2} \frac{1-y} {A(y)} \frac{x f_{+3dL}^{\perp} } {f_1}, \label{eq:Asin2ul} \\
&\langle{\sin (\phi+\phi_S)}\rangle_{UT}= 
 - x\kappa_M \frac{ |\vec k_\perp|^3}{M^3} \frac{1-y}{A(y)} \Bigl(\frac{f_{+3dT}^{\perp 4}}{f_1} + \frac{f_{-3dT}^{\perp 2}}{f_1}\Bigr), \label{eq:Asinut2}\\
&\langle{\sin (3\phi-\phi_S)} \rangle_{UT}= 
 - x\kappa_M \frac{ |\vec k_\perp|^3}{M^3} \frac{1-y}{A(y)} \Bigl(\frac{f_{+3dT}^{\perp 4}}{f_1} - \frac{f_{-3dT}^{\perp 2}}{f_1}\Bigr). \label{eq:Asinut3}
\end{align}
They have only twist-4 contributions up to this level and can therefore serve as good places to study such twist-4 effects.
 
It is clear that if we insert the relationships given by Eqs.~(\ref{eq:f3dg0}-\ref{eq:g3ddTperpg0}) 
into Eqs.~(\ref{eq:Asin1ut}-\ref{eq:Asinut3}) , we obtain that results for $g=0$ 
such as those obtained in \cite{Cahn:1978se,Liang:1993re}. 
The deviations from them reflect the effects of multiple gluon scattering. 

We note in particular that by replacing $\phi$ by $\phi_h$, 
the 6 azimuthal asymmetries given by Eqs.~(\ref{eq:Asin1ut}-\ref{eq:Acoslt2}) and (\ref{eq:Acos2uu}-\ref{eq:Asinut3}) 
are just the 6 twist-2 asymmetries in $e^-N\to e^-hX$. 
Measurements of them are one of the major tools that we use to extract twist-2 TMDs. 
Here we see clearly that, even if the fragmentation part is not considered, there are twist-4 contributions to all of them. 
We emphasize that the factor $(1-\alpha_{UU}\kappa_M)$ is due to the twist-4 contributions to $W_{UU}$. 
It exists for all azimuthal asymmetries that have twist-2 contributions. 
This means that this is the least modification factor that we have for all the 6 twist-2 azimuthal asymmetries for $e^-N\to e^-hX$.  

In view of that $Q^2$ from experiments such as HERMES or JLab 
(see e.g. \cite{Airapetian:2009ae,Avakian:2010ae}) are usually from 1 to $10$GeV$^2$ so $\kappa_M$ takes values from $0.1$ to $1$, 
the twist-4 modifications can be quite large depending on the coefficient of $\kappa_M$ in equations given above. 
A reliable estimation of these twist-4 contributions depend on the unknown twist-4 PDFs involved. 
We note that there are totally 18 independent twist-4 TMD PDFs involved in the final results, 
2 from $\hat \Phi^{(0)}$, 4 pairs from  $\hat\varphi^{(1)}$,  and 4 pairs from $\hat\varphi^{(2)}$. 
These twist-4 TMDs contain information on intrinsic parton distribution in nucleon and effects of multiple gluon scattering contained in the gauge link. 
They contain in particular quantum inference effects in the multiple gluon scattering thus there are no simple probability interpretations. 
Clearly, there is a long way to go to make precise measurements of all of them. 
Presently, being lack of knowledge about these twist-4 TMD PDFs, as a crude approximation, 
we suggest to use relationships between the higher twist and the twist-2 TMDs obtained at $g=0$ given by Eqs.~(\ref{eq:f3dg0}-\ref{eq:f3ddTperpg0}) 
and assume that they are approximately valid also at $g\not=0$ to make rough estimations of twist-4 effects. 

In this case, we obtain that $\langle\sin2\phi\rangle_{UL} \approx 0$,  the other three twist-4 asymmetries become,  
\begin{align}
&\langle\cos2\phi\rangle_{UU} \approx  2\kappa_M \frac{|\vec k_\perp|^2}{M^2} \frac{1-y} {A(y)}, \label{eq:aAcos2uu} \\
&\langle{\sin (\phi+\phi_S)}\rangle_{UT}\approx  - \kappa_M \frac{|\vec k_\perp|^3}{M^3}    \frac{1-y}{A(y)}   \frac{f_{1T}^\perp}{f_1}, \label{eq:aAsinut2}\\
&\langle{\sin (3\phi-\phi_S)} \rangle_{UT}\approx  \kappa_M \frac{|\vec k_\perp|^3}{M^3}  \frac{1-y}{A(y)}    \frac{f_{1T}^\perp}{f_1}, \label{eq:aAsinut3}
\end{align}
and the modification factors for the two twist-2 asymmetries given by Eqs.~(\ref{eq:Asin1ut}-\ref{eq:Acoslt2}) become, 
\begin{align}
& 
   \alpha_{UT} \approx \frac{|\vec k_\perp|^2}{M^2}\Bigl[
   - \frac{\partial \ln f_1}{\partial \ln x}
   + \frac{\partial \ln f_{1T}^\perp}{\partial \ln x} \Bigr], \label{eq:aalphaut} \\
&
   \alpha_{LT} \approx \frac{|\vec k_\perp|^2}{M^2}\Bigl[ \frac{8(1-y)}{A(y)}
   - \frac{\partial \ln f_1}{\partial \ln x} 
   - \frac{\partial \ln g_{1T}^\perp}{\partial \ln x} \Bigr], \label{eq:aalphalt}\\
& \alpha_{UU} \approx \frac{|\vec k_\perp|^2}{M^2} \Bigl[ \frac{8(1-y)}{A(y)} - \frac{\partial \ln f_1}{\partial \ln x} \Bigr], \label{eq:aalphauu}
\end{align}
where we see that the first term in the square bracket for $\alpha_{LT}$ or $\alpha_{UU}$ can already reach 4 
and the second term is additive to it. This shows that the twist-4 contributions could indeed be very significant.
At the present stage, Eqs.~(\ref{eq:aAcos2uu}- \ref{eq:aalphauu}) could serve as a rough estimation of them in SIDIS.

{\it{Summary}} --- 
In summary, 
benefited from the collinear expansion, we carried out the calculations up to twist-4 and 
present for the first time the complete twist-4 result for $e^-N\to e^-qX$ with polarized beam and target.
The results show that, among the 18 structure functions, besides the 8 that have only twist-3 contributions, 
all the other 10 have twist-4 contributions. 
We show in particular that among these twist-4 contributions, 4 correspond to azimuthal asymmetries 
where twist-4 are the leading power contributions in  $e^-N\to e^-qX$ and can serve as good places to study these twist-4 effects. 
We show also that for all the 8 twist-2 structure functions for $e^-N\to e^-hX$ that correspond to the 8 twist-2 TMD PDFs, 
there are twist-4 addenda to them. 
These twist-4 contributions could be quite significant and have strong impact on the study of TMD PDFs 
in particular in the energy regions of existing DIS experiments such as HERMES and those in JLab. 
We suggest an approximate way for rough estimations of twist-4 contributions using corresponding twist-2 PDFs.

{\it{Acknowledgements}} ---
This work was supported in part by the Major State Basic Research Development Program in China (No. 2014CB845406), 
the National Natural Science Foundation of China (Nos. 11375104 and 11675092),  
and the CAS Center for Excellence in Particle Physics (CCEPP).

\end{document}